\documentclass[12pt,a4paper]{article}
\usepackage{natbib}
\usepackage{geometry}
\usepackage[latin1]{inputenc}
\usepackage{graphicx,graphics,textcomp}%varioref,bookman
\usepackage{fancyhdr}
\usepackage{lscape,epsfig,amssymb,amsmath}
\usepackage[]{subfigure}
\usepackage{rotating,placeins}
\usepackage{lscape}
\usepackage{setspace}
\newenvironment{myindentpar}[1]%
  {\begin{list}{}%
     {\setlength{\leftmargin}{#1}}%
     \item[]%
  }
  {\end{list}}

\title{The International Workshop on Wave Hindcasting and Forecasting and the Coastal Hazards Symposium}
\author{{\O}yvind Breivik\footnote{Final version published in
\textit{Ocean Dyn}, 2015, doi:10.1007/s10236-015-0827-9.}\thanks{Corresponding author. E-mail:
\texttt{oyvind.breivik@met.no}; ORCID Author ID: \texttt{0000-0002-2900-8458}.
Norwegian Meteorological Institute, Alleg 70, NO-5007 Bergen, Norway}, 
\and Val Swail\thanks{Climate Research Division, Environment Canada, Toronto, Canada},
\and Alexander V Babanin\thanks{Swinburne University of Technology, Australia},
\and Kevin Horsburgh\thanks{National Oceanography Centre, Liverpool, UK}
}

\begin{document}
\maketitle                                                                                                          

\begin{abstract}
Following the 13th International Workshop on Wave Hindcasting and Forecasting
and 4th Coastal Hazards Symposium in October 2013 in Banff, Canada, a topical
collection has appeared in recent issues of \emph{Ocean Dynamics}.
Here we give a brief overview of the history of the conference since its
inception in 1986 and of the progress made in the fields of wind-generated
ocean waves and the modelling of coastal hazards before we summarize the main
results of the papers that have appeared in the topical collection.

\end{abstract}

\section{History of the Workshop}
The first International Workshop on Wave Hindcasting and Forecasting was
held in Halifax, Nova Scotia, Canada in September 1986. Since then, 12 more
workshops have been organized, joined in 2007 by the Coastal Hazards Symposium.

Wave forecasting and its close relative wave hindcasting (distinguished from
reanalyses by the absence of assimilated data) have undergone dramatic changes
since the early days of the workshop. Back then the now common third generation
(3-G) wave models where the model spectrum evolves freely through nonlinear
interaction had only recently been developed \citep{swamp85,has88}. The workshop
intended to bring hands-on users of wave data in contact with the leading
theoreticians and modellers of the day. In this it succeeded, and a number of
important projects were born out of contacts established at the conference where
the oil and gas producers in particular could specify what their needs were. As
such the workshop helped anchor the wave modellers, usually people with a
penchant for theoretical musings, to practical problems facing users who had to
weather storms in some of the windiest places on earth.

Although the precise nature of the research has changed significantly over
the past 28 years, the primary objectives of the workshop have remained
essentially the same, namely to
\begin{itemize}
  \item provide a forum for the exchange of ideas and information related to
  wind and wave hindcasting and forecasting, including modelling, measurement
  and past and future states of the climate
  \item coordinate ongoing research and development initiatives
  \item discuss priorities for future research and development
\end{itemize}
The period around the time of the initial workshop was a very exciting one for
wave modelling, forecasting and hindcasting. The WAM model \citep{has88} was in
the process of becoming the first community wave model, arising out of the SWAMP
program \citep{swamp85} and the plethora of wave models which existed in the
early 1980s. The advent of these sophisticated third-generation models, and the
rapidly increasing computing capabilities of numerical weather prediction
centres, led to the increased use of numerical wave models in operational
forecasting.

With time more centres began using these models for their operational
programs, and an international wave forecast comparison project was
established, coordinated by The European Centre for Medium-Range Weather
Forecasts (ECMWF), to evaluate forecast performance and identify areas of
potential improvement \citep{bidlot02}. At the same time, hindcasting was
becoming the accepted basis for developing wave climatology in general,
and particularly for developing the basis for offshore design criteria.
Due to the limitations of computing power and the lack of suitable historical
wind fields, these early hindcasts covered only the most extreme storms in a
limited area of the ocean.  They were thus a far cry from today's decades-long,
global wave hindcasts forced with atmospheric reanalyses. Yet at the time
these hindcasts took a Herculean effort in terms of computing and through
the meticulous preparation of suitable wind fields.

An important alliance was formed in 2006, when the workshop was first
co-sponsored by the WMO-IOC Joint Technical Commission for Oceanography and
Marine Meteorology (JCOMM). Subsequently, and following the very successful
JCOMM Scientific and Technical Symposium on Storm Surges in October 2007 in
South Korea, a symposium on hazard assessment in coastal areas was established,
and later held jointly with the Wave Workshop.  This has provided
an excellent opportunity for cross-pollination of ideas between wave and
storm surge prediction.  While the primary focus of the workshop was, and
remains, wave hindcasting and forecasting, what makes this workshop unique
is the treatment of the end-to-end research issues associated with ocean
waves, from the basic research on wave physics to the ultimate use of the
products. The topics have thus ranged from research and operational aspects of
wave and storm surge hindcasting and forecasting, regional hindcasts, storm
surge climatology, data collection and instrumentation, data assimilation,
wave-current interaction, wave-ice interaction, shallow-water and nearshore
effects, wind fields for wave hindcasting or forecasting, extremal analysis,
as well as past and future climate trends and variability.

Over the past 30 years the Wave Workshop has presented the results of
many innovative and groundbreaking studies, introduced new national and
international initiatives, and induced a large number of collaborative research
efforts on all aspects of wave research, including measurements, modelling,
forecasting and validation, as well as wave climate including historical
trends and future projections. It would be impossible to list all of those
here, and we would risk omitting many important interactions. However,
a few noteworthy examples are described in the following paragraphs.  On a
popular note, the 10th Wave Workshop is well described in Chapter 8 of the
popular-science book \textit{The Wave} by \citet{casey10}.  We also remind
the reader that the proceedings of all workshops are available online at
\underline{\texttt{http://www.waveworkshop.org}}.

The initial Wave Workshop (WW-1) had three particularly noteworthy
presentations. \citet{komen86} described the development of the landmark WAM
model \citep{has88}, the first 3-G spectral wave model. The model resulted
from a major international collaboration led by Klaus Hasselmann following
the Sea Wave Modelling Project \citep{swamp85} which had demonstrated large
discrepancies in the fauna of wave models in use at the time. Its
presentation at WW-1 served to broaden the community even further.  WW-1 also
saw two landmark presentations related to wave hindcasting and its use for
structural design.  \citet{francis86} described the North European Storm Study
(NESS), which was the first effort to construct a long-term climatology of
extreme wave events based on selected storms for an ocean basin. At the same
time, Sverre Haver reported on the adequacy of hindcast data for structural
design, illustrating the industry's move towards hindcasting as the basis
for design criteria instead of relying solely on measurements.

The second Wave Workshop (WW-2) boasted an entire session
\citep{cardone89a,cardone89b,szabo89a,szabo89b,bolen89} on wave hindcasting
of extreme storms for the development and evaluation of the wave criteria
for the design of the gravity-based production platform at Hibernia, on the
Grand Banks of Newfoundland. This was the first time that the industry's
design procedures had been published in such a forum.

WW-3 saw the introduction of data assimilation methods in wind wave modelling
\citep{burgers92} and a coupled wind-wave data assimilation system by
\citet{heras92}. WW-3 also saw a presentation of ``Hindcasting waves using
a coupled wave-tide-surge model'' by \citet{wu92}, which would be an early
precursor to work done in later years on coastal inundation, incorporating
wave, tide and surge components.

Two significant results were reported at WW-4. The presentation ``The
WASA project: changing storm and wave climate in the northeast Atlantic and
adjacent seas?'' \citep{was97a,was97b} described a continuous hindcast archive
as well as a hindcast of selected storms forced with archived operational
wind fields for the northeast Atlantic, and a comprehensive climate trend
analysis. \citet{cox95} introduced ``An interactive objective kinematic
analysis system'' as an efficient way to incorporate manual kinematic analysis
of surface wind fields into background fields from numerical weather prediction
products. This technique yields a much better representation of the highest
winds in a storm, and thus also of the most extreme wave conditions which drive
design criteria. This approach has been refined over time, and remains the most
effective method to capture the highest sea states in a hindcast, even with
the advent of high quality reanalyses as evaluated later by \citet{caires04}.

WW-5 was very noteworthy for the introduction of the first continuous wind-wave
hindcast forced with reanalysis winds.  \citet{cox01} and \citet{swail00}
respectively described the evaluation of the marine surface wind product
from the National Center for Environmental Prediction (NCEP) reanalysis
\citep{kalnay96} and the application of those forcing fields, including the
kinematic wind analysis previously developed, to produce a long-term North
Atlantic wave hindcast. On the wave modelling side, WW-5 saw the presentation
of the third-generation SWAN wave model \citep{boo99,ris99}. The end users
continued to be well represented in this forum, as illustrated by the
presentation on ``Wave Hindcasting and Forecasting: Their Role in Ensuring
the Safety of Personnel Involved in Offhsore Oil and Gas Exploitation''
by \citet{smith98}.

WW-6 introduced the rapidly expanding research topic of non-linear wave
phenomena, particularly with respect to rogue or freak waves described in
the presentation ``Nonlinear dynamics of rogue waves'' by \citet{osborne00}.

WW-7 introduced several new concepts. \citet{swail02} showed the first attempts
to describe not only the past trends of the wave climate from a reanalysis,
but also future projections based on statistical downscaling methods for
an entire ocean basin, the North Atlantic.  \citet{thomas02,thomas05}
demonstrated the importance of homogenisation of historical wave
measurements for removing spurious trends due to changes in hull type,
sensor and processing methods. Also, \citet{gulev02} demonstrated that,
while visual wave observations had long been considered to be a barely
adequate estimate of the wave climatology on a global or regional basis,
it was still possible to derive more sophisticated information from them
(see also \citealt{gulev03,gulev04}). Following up on some of the research
on rogue waves, \citet{gunson02} presented an approach for investigating
conditions for rogue wave events using a global spectral wave model. Not
least among the key presentations at WW-7 was the introduction of the 2002
release of WaveWatch-III \citep{tolman02}.

WW-8 saw the wave climate trend and future projection research advance to the
identification of climate change signal and uncertainty \citep{wang04}. The
usefulness of satellite wind fields for wave hindcasting, in particular from
QuikScat, was highlighted by \citet{cardone04}. Scatterometers in general,
and QuikScat in particular, have proven to be an unparalleled resource for
both wave hindcasting and forecasting. Another key area of development in
the modelling realm was described by \citet{banner04}, on forecasting of
breaking waves during storms.

WW-9 described the first ever detection of human influence on
trends of North Atlantic Ocean wave heights and atmospheric storminess
\citep{wang06a,wang06b}. This workshop also saw a focus on extreme waves, with
the presentation on extreme waves in the ECMWF operational wave forecasting
system \citep{bidlot06}. \citet{grigorieva06} also extended their previous
work on wave observations from voluntary observing ships (VOS) to show their
utility for extreme waves worldwide and their changes over the past 50 years.

At WW-10, \citet{bidlot07b} and \citet{hines07} described advances in the
JCOMM Wave Forecast Verification Project, a routine inter-comparison of wave
model forecast verification data to provide a mechanism for benchmarking
and assuring the quality of wave forecast model products, which had been
formalized as a JCOMM activity at the first session of the JCOMM Expert
Team on Waves and Surges in 2003, and was subsequently described in a JCOMM
Technical Report by \citet{bidlot06b}. The project now includes 17 centres,
many running global wave forecast systems, with different wave models,
different wind forcing, and different model configurations. The goal is
to continue to add new participants, including regional participants (where
appropriate), and to expand the scope of the intercomparison as feasible. This
includes validation against satellite altimeter observations but also comparing
spectra as well as utilizing techniques for comparing against measurements
spatially separated from the model grid point. Here it is pertinent to mention
in particular the triple-collocation technique pioneered by \citet{caires03}
and later used by \citet{janssen07} and \citet{abdalla11} for random error
estimation in altimeter wind and wave products.  Extreme wave prediction was
also a focus of this workshop, with advances in freak wave prediction from
spectra described by \citet{mori07}, and the extension of the ECMWF freak
wave warning system to two-dimensional propagation \citep{wam40r1}.

WW-11 saw advances in the understanding of wave measurements.  \citet{swail09}
described a philosophy for continuous wave measurement and evaluation as
presented to the Oceanobs'09 conference, and the development of a JCOMM
Pilot Project on Wave measurement Evaluation and Test (WET). \citet{bender09}
presented evidence showing very large potential overestimates of wave height
measurements in buoys with a simple 1-D accelerometer due to heeling of
the hull. Mark Hemer introduced the joint JCOMM and World Climate Research
Programme (WCRP) Coordinated Ocean Wave Climate Projections (COWCLIP)
initiative, which is looking at potential future changes to the wave
climate (see \citealt{hemer12} for an overview of COWCLIP).  On the wave
forecasting side, \citet{banner09} described new methods for incorporating
breaking wave predictions into spectral wave forecasting models (see also
\citealt{banner10}). An entire session was devoted to the emerging forecast
activity of ensemble wave predictions, including those of the MetOffice,
ECMWF, Meteo-France, NCEP and Environment Canada.

At WW-12 the recently developed NCEP Climate Forecast System Reanalysis (CFSR)
was presented \citep{chawla13}.  This data set entails a coupled reanalysis
of the atmospheric, oceanic, sea-ice and land data from 1979 through 2010,
and a reforecast run with this reanalysis \citep{saha10}. This reanalysis
has much higher horizontal and vertical resolution than the Global and the
North American Reanalysis.  In keeping with the end-to-end scope of the
workshop, \citet{carrasco11} investigated the potential for ensemble wave
forecasts of weather windows with the requirements of offshore operations
in mind. In particular, the issue was addressed as to whether the emerging
probabilistic forecast approach is (always) better than a deterministic
one, and whether probabilistic decisions are best based on ensemble mean or
probability threshold.

\section{The 13th Workshop}
The 13th International Workshop on Wave Hindcasting and Forecasting and the
4th Coastal Hazards Symposium was held in Banff, Canada from 27 October to
1 November, 2013. The workshop had more than 90 participants who gave over
75 presentations and displayed 18 posters.

In addition to the selection of papers from WW-13 submitted for this special
issue of Ocean Dynamics, there were several keynote presentations. In view
of the recent increased interest in the Arctic, two papers were particularly
noteworthy. \citet{bidlot14} described a new approach to inclusion of sea
ice attenuation in an operational wave model (as opposed to the standard
operational approach at ECMWF to consider 30\% ice concentration and above
as land and spectra set to zero, see also \citealt{doble13}). As well,
\citet{salisbury14} described the implications of ice cover on storm
surge dynamics in the Beaufort Sea. An important new initiative of WMO,
the Coastal Inundation Forecasting Demonstration Project (CIFDP), combining
flood forecasting due to storm surge, ocean waves, tides and river runoff,
was introduced by \citet{lee14}.

In addition to the general sessions, each workshop has had a theme session. For
WW-13 this theme session targeted the JCOMM priority area ``Forecasting
Dangerous Sea States''. Topics included theoretical, numerical, laboratory or
operational applications dealing with forecasting of hazardous wave conditions
such as crossing seas, unusually steep waves, rapidly developing seas and
rogue waves. In addition to the main topic the sessions spanned a wide range
of topics, among those wave measurements, operational wave and storm surge
forecasts, hindcasting, advances in wave and storm surge modelling, coastal
impact and a special session on COWCLIP (see \citealt{hemer12,hemer13}).

% Session C: Forecasting Dangerous Sea States
Forecasting dangerous sea states is a diverse topic since what is dangerous
to one vessel or platform may be harmless to another.  This was reflected by
the large number of presentations ranging from studies of the impact of rogue
waves on marine structures \citep{bitner14b} to theoretical investigations
into the generation of extreme waves from interactions between fully
nonlinear 2D surface waves \citep{chalikov14}. A similar model was used by
\citet{babanin14} to investigate the interaction of wave components that
are very close in wavenumber space, and it was found that the interactions
between wave components seemingly break down when the separation in wavenumber
space becomes very small. This was taken as evidence of a ``corpuscular''
nature of water waves, i.e., that there is an inherent discreteness to the
Fourier spectrum of waves.

In situ measurement of waves remains the main source of ``ground truth'' for
assessing the quality of wave models, both for hindcasting and for operational
forecasts. The vexing issue of what exactly buoys measure, especially in
high seas, remains a challenge some 60 years after the first wave spectra
were measured using capacitance wire recorders \citep{tucker55,burling55}.
\citet{waseda14} demonstrated that buoys originally meant for meteorological
measurements or tsunami monitoring can be equipped with GPS sensors and provide
reliable wave measurements in remote deep-water locations. Observations
from tropical cyclones showed evidence of nonlinearities in the orbital
motion of the highest waves in the centre of wave groups. This is of
interest when studying the behaviour of freak waves.  \citet{collins14}
remind us that wave buoys normally report only integrated parameters. This
is insufficient for investigating the extremal behaviour of individual waves
(rogue waves) but understandable in light of the low bandwidth and limited
storage of traditional wave measuring devices in remote locations. However,
\citet{collins14} demonstrate that by also reporting the maximum wave height
and the total number of waves per record, valuable information about the
occurrence of extreme waves can be gleaned from properly tilt-corrected
wave buoys.

\citet{bitner14} investigated the intrinsic sampling variability of the most
frequently used wave-height and wave-period scales, namely significant wave
height and zero-crossing wave period, by varying the recording intervals of
2~Hz time series from a wave buoy and a downward looking laser in the central
North Sea.  It is argued that sampling variability from wave measurements
is greater than from models because of natural long-term trends in the field
and due to non-stationary changes of weather conditions. Standard deviations
of significant wave height and zero-crossing wave period are compared with
observations and show similar trends. The sampling variability is higher in
wave height than in wave period, but both increase as the height and period
grow. It is also noted that the JONSWAP spectrum \citep{has73} gives higher
variability than the Pierson-Moskowitz spectrum \citep{pie64}.

Remote sensing of ocean waves by microwave radar is a relatively
well-understood process where Bragg scattering from short gravity-capillary
waves, which in turn are modulated by the presence of longer gravity waves,
allow a spectrum to be estimated \citep{wri68,kel75,phi88}. Although the
relative energy of the ocean wave spectrum may be estimated quite precisely,
establishing the absolute energy level may be a harder task. \citet{ewans14}
have carefully analyzed the principles behind the measurement technique and
long time series of observations from radars operated by Shell.  The conclusion
is that the radars compare remarkably well to Wavec buoys in the vicinity of
three platforms in the North Sea as well as one in the Norwegian Sea. The
observations are stratified by direction to analyze the shielding from the
platform, but only small reductions are found in directions where shielding
may be expected. \citet{gibson14} investigated field measurements of wave
height and crest elevation from Saab WaveRadar instruments mounted on eight
fixed jacket platforms and a Datawell Directional Waverider buoy during a
severe storm in the North Sea. The storm generated seas which peaked well in
excess of the 100-year significant wave height for that region. The authors
also identify 19 freak-wave events. They found that the significant steepness
and spectral bandwidth during the storm remain almost constant. Consequently,
there is little change in the commonly applied design wave height and crest
elevation probability distributions during the storm. Whilst the bulk of the
recorded data was in good agreement with the theoretical distributions,
it was demonstrated that when the wind speed exceeded 25~m~s$^{-1}$,
the measured crest elevation lay above the second-order Forristall wave
crest distribution \citep{forristall00}. The authors postulate that this
could be due to the modulational instability of the waves due to both the
wind forcing and/or their steepness or local wind input  to large waves.
\citet{lund14} noted that the strength of the surface wave signal in marine
X-band radar images strongly depends on range and azimuth,  i.e. the angle
between antenna look and peak wave direction. The field of view is typically
partially obstructed, e.g. due to the coastline or ship superstructures, which
may result in an increased variability or error associated with estimated
wave parameters. Using reference data from a nearby Datawell Waverider
buoy off the California coast, the authors quantified the dependency of the
radar-based two-dimensional wave spectrum and parameters on range and azimuth
and proposed and evaluated empirical methods to remove the dependency.
In a related study \citet{hessner14} demonstrated the large amount of
information that can be gathered in coastal locations through the use of
high resolution remote sensing techniques. High resolution wave, current
and water depth fields derived by marine X-band radar were presented for a
coastal region of extreme tidal currents in the presence of inhomogeneous
bathymetry at the south coast of New Zealand's North Island. The sea state
data provides a spatial representation of coastal effects like wave shoaling
and refraction forced by bathymetry and current interaction. The near-surface
current measurements about 3 km off the coast show expected tidal current
pattern, in agreement with currents from RiCOM, a hydrodynamic model. The
observed current field also captured small-scale features caused by the
influence of the local bathymetry.

% Session D: Advances in Storm Surge Modelling
The modelling and forecasting of storm surges was the subject of three sessions.
An emerging theme is probabilistic forecasting of water level using ensembles of
wind and pressure fields. A number of operational ensemble forecast systems are
already in place around the world
\citep{flowerdew09,devries09,flowerdew12,flowerdew13}. \citet{mel14} presented
results from an ensemble for the Adriatic Sea forced by ECMWF wind and pressure
fields. In a three-month period in 2010 the model was found to have a good
spread-error ratio and the ensemble mean is found to outperform deterministic
forecasts. Hence the ensemble serves both to forecast the uncertainty as well as
improve the central forecast.  \citet{etala15} report advances in storm surge
forecasting for the Argentian coast.  Using a Local Ensemble Transform Kalman
Filter (LETKF), water level from in situ gauges and satellite altimetry is
assimilated into a high-resolution 20-member ensemble storm surge model. The
improvement on the inner part of the shelf is significant, and demonstrates
quite clearly the value of using satellite altimetry for storm surge forecast
systems.

% Session G: Design
The design criteria of offshore installations are critically sensitive to the
highest waves encountered in a storm.  \citet{bitner14b} investigated the
occurrence of rogue sea states and the consequences for marine structures in a
session on design criteria.  They looked in detail at crossing sea states where
rogue waves are likely to occur. This is done on the basis of a hindcast with
the the WAM model, combined with numerical simulations in the physical space by
means of the Higher Order Spectral Method. It is concluded that the 40-degree
separation of two wave systems is where the largest waves can be expected.

The 100-year return values of significant wave height and 10-m wind speed are
important parameters for the estimation of the load and fatigue on offshore
installations. A new method for estimating return values from very large
ensembles of archived forecasts at long lead times was presented by
\citet{bre14b}. Nine years of archived 10-day forecasts consisting of 51 members
from the ECMWF ensemble were used to produce global maps of return values for
10-m wind speed and significant wave height. The method was previously used to
explore wave height return values in the North Atlantic \citep{bre13b}.

% Session H: Operational Wave Forecasting
The session on operational wave forecasting saw the public release of the fourth
WAVEWATCH III version (the model development was previously described in WW-12,
see also \citealt{tolman13}). One of the contributions to the new WAVEWATCH III
version is the option of running on a spherical multiple-cell grid which allows
regions of various resolution to be handled in a single model integration.
\citet{li14} present a global implementation at resolutions of 25, 12 and 6 km
and find that the performance compares favourably with the UK Met Office
regional wave forecasts.

% Trends Session M: Wave Climate Trend and Variability
The session on wave climate trend and variability saw a number of papers
exploring new methods to tease out the trends of the past and future wave
climate. \citet{aar15} presented a detailed analysis of the trends in wave
height and marine wind speed from ERA-Interim \citep{dee11}. They found that the
trend at analysis time is contaminated by changes to the observation network.
This is particularly problematic for wave height since the wave model relies
solely on altimeter observations. There is an abrupt change in the mean wave
height in 1991 when the ERS-1 satellite was launched.  10-day forecasts starting
at 00 and 12 UTC are part of the ERA-Interime reanalysis runs, and it was found
that a lead time of 48 hours gave a more realistic trend in both wave height and
wind speed.  \citet{vanem14} used a Bayesian spatiotemporal analysis on the
NORA10 hindcast \citep{rei11,aar12,bre13b} to investigate the trends present in
the hindcast. The trends were found to be weaker than those found using a
similar method on the C-ERA-40 data set \citep{cai05}, which is a statistical
adjustment of the ERA-40 reanalysis \citep{upp05}. \citet{paris14} compared the
trends in wave height for the Bay of Biscay from a number of coarse resolution
reanalyses and hindcasts to their recently completed high-resolution hindcast
BOBWA-10kH \citep{charles12}. They find that the regional forecast slightly
overestimates the wave height in the Bay of Biscay whilst ERA-Interim and
C-ERA-40 tends to underestimate slightly. The trends appear to be broadly
similar.  In a similar study, \citet{grabemann14} used a consistent approach to
analyze a set of ten wave climate projections to estimate the possible impact of
anthropogenic climate change on mean and extreme wave conditions in the North
Sea (see also \citealt{groll14}).  They found that though the spatial patterns
and the magnitude of the climate change signals vary, some robust features among
the ten projections emerge: mean and severe wave heights tend to increase in the
eastern parts of the North Sea towards the end of the twenty-first century in
most projections, but the magnitude of the increase in extreme waves varies by
several decimeters between the projections. For the western parts of the North
Sea more than half of the projections suggest a decrease in mean and extreme
wave heights. They conclude that the influence of the emission scenario on the
climate change signal seems to be less important, and that the transient
projections show strong multi-decadal fluctuations, and changes towards the end
of the twenty-first century might partly be associated with internal variability
rather than with systematic changes.

\citet{perez14} presented a diagnostic method for analyzing the origin and
travel time of wave systems for a given location, known as ``Evaluation of
Source and Travel-time of wave Energy reaching a Local Area'' (ESTELA).  The
method allows the characterization of the area of influence for a particular
location.  The method is capable of characterizing the wave energy and travel
time in that area and relies on on a global scale analysis using both
geographically and physically based criteria. The geographic criteria rely on
the assumption that deep water waves travel along great circle paths. This
limits the area of influence by neglecting energy that cannot reach a target
point, as its path is blocked by land.  In a companion paper, \citet{camus14}
described a method for identifying optimal predictors when investigating which
indices to use for assessing the local wave climate.  The method is based on a
statistical model that relates significant wave height to the sea level
pressure.  The predictor is composed of a local and a regional part,
representing windsea and swell, respectively. The spatial domain of the
predictor is determined using the ESTELA technique.

% Session N: Wave Hindcasts
Several regional hindcasts have appeared recently, and the session on wave
hindcasts at this workshop saw the presentation of a truly high-resolution
hindcast for Irish waters.  \citet{gallagher14} employed an unstructured grid
varying in resolution from about 250~m nearshore to about 10~km offshore with
ERA-Interim as boundary conditions.  \citet{semedo15} investigated the swell
climate in the Nordic Seas as it is represented by the NORA10 hindcast
\citep{rei11}. They found that swell is dominant more than 50\% of the time even
in the relatively sheltered Barents Sea during the winter months. During summer
months, the swell prevalence rises to about 90\% in the Norwegian Sea. Only in
the semi-enclosed North Sea did the winter-time swell prevalence drop below
50\%.  Large changes in the wintertime atmospheric circulation over the North
Atlantic during the period had an impact on the wave heights in the eastern
North Atlantic mid to high latitudes.  Decadal trends of total significant wave
height in the Nordic Seas are mostly due to contribution of swell and to the
changes in wave propagation.  \citet{carrasco14} investigated the global wave
drift climate, i.e., the geographical distribution of the surface Stokes drift
velocity and the Stokes mass transport, using ERA-40 \citep{upp05}. The study
shows that in most of the oceanic basins the global surface Stokes drift is
chiefly driven by the local wind sea while its vertically integrated transport
is mainly swell-driven.  \citet{poncedeleon15} calculated the Benjamin-Feir
index \citep{janssen03} from a wave model integration forced with winds from the
CFSR reanalysis \citep{saha10} to investigate the presence of extreme sea states
in North Atlantic extratropical storms.

% COWCLIP session
The COWCLIP session on the future wave climate was the second of its kind.
There is still a debate about the cost-benefit ratio of adding dynamical wave
models to future coupled model intercomparison project climate integrations
(CMIP).  In the meantime, a few estimates of the impact of waves on the climate
system are starting to appear. In this session \citet{fan14} presented coupled
climate simulations of the impact of wave-induced Langmuir turbulence
\citep{sky95,mcwilliams97,teixeira02,polton07,grant09,belcher12} and turbulence
from non-breaking waves \citep{qiao04,babanin09,qiao10}. This is an important
topic for both climate modelling as well as coupled models on seasonal and
shorter scales as it has implications for how the ocean surface boundary layer
(OSBL) warms.

The 22 articles in the topical collection provide a snapshot more than a
complete overview of the current state of the field. We hope that by putting
together this topical collection we provide a starting point for new workers in
the field as well as a body of references of what has been published earlier.
The Wave Workshop will continue its bi-annual progress, and exciting times lie
ahead as wave modellers and ocean modellers take the first tentative steps
toward each other by introducing effects of the wave field on the ocean surface
boundary layer in full-fledged coupled models of the atmosphere-wave-ocean
system \citep{babanin09b,janssen13,fan14,bre15tmp}. The effect of the sea state
on the climate system is still relatively uncharted territory, but it appears
increasingly likely that the enhanced mixing through wave breaking
\citep{craig94} and Langmuir turbulence \citep{mcwilliams97,grant09,belcher12}
together with the Coriolis-Stokes forcing
\citep{hasselmann70,weber83,jenkins87b,mcwilliams99,polton05,bre14} have a
profound effect on the mixing in the upper ocean. This in turn has a direct
feedback effect on the climate system through the heat fluxes between ocean and
atmosphere. But there is also another way through which waves affect the climate
system, namely the increased (or reduced) roughness on the sea surface. This was
explored by \citet{janssen89,janssen91} and later implemented in the coupled
atmosphere-wave forecast system at ECMWF in 1998 \citep{jan04}. The increased
roughness affects the evolution of low pressures in the extratropics, but there
is also evidence of the opposite in swell-dominated areas \citep{semedo09}.
Finally, the role of waves in the marginal ice zone is beginning to be better
understood. \citet{doble13} and \citet{kohout14} show the dramatic reach of
storms in ice-covered waters through measurements of wave activity more than
100~km from the ice edge. Waves probably play an important role in breaking up
the ice in the marginal ice zone, causing the ice to melt more rapidly. But
waves are also affected by the presence of sea ice as its presence changes the
fetch, and it seems clear that only coupled models incorporating both an active
ice model with wave break up and a wave model which can model the sea state in
areas with moderately high sea ice concentration can hope to realistically
capture this interaction. We look forward to seeing presentations on these and
other topics relating to the modelling of ocean waves in future workshops, the
first of which is to be held in Key West, Florida, in November 2015.

\section*{Acknowledgments}
The conference co-chairs would like to express their gratitude to the
organizers and sponsors: JCOMM, Environment Canada and the US Army Corps of
Engineers. More information about the conference can be found at
\begin{myindentpar}{1.5cm}
  \underline{\texttt{http://www.waveworkshop.org}}
\end{myindentpar}
We are grateful to \emph{Ocean Dynamics} for taking the subject into
consideration for a topical collection. {\O}B has benefited from the European
Union FP7 project MyWave (grant FP7-SPACE-2011-284455) and the ExWaCli project,
funded by the Research Council of Norway (grant no 226239), while editing this
topical collection.

%\bibliography{../../Breivik}

\newpage

\end{document}